\pdfoutput=1

\documentclass[supp]{new_tlp}
\usepackage{supp}
\usepackage{mathptmx}

\usepackage[pdftex]{graphicx}
\usepackage[pdftex]{hyperref}
\usepackage{amsmath}
\usepackage{amssymb}
\usepackage{esvect}
\usepackage{multirow}

\newenvironment{changemargin}[2]{%
\list{}{\rightmargin#2\leftmargin#1
\parsep=0pt\topsep=0pt\partopsep=0pt}
\item[]}
{\endlist}

\newenvironment{indented}{\begin{changemargin}{1cm}{0cm}}{\end{changemargin}}

\newenvironment{packedenum}
{\begin{enumerate}
  \setlength{\itemsep}{3pt}
  \setlength{\parskip}{0pt}
  \setlength{\parsep}{0pt}
}
{\end{enumerate}}

\newenvironment{packeditems}
{\begin{itemize}
  \setlength{\itemsep}{3pt}
  \setlength{\parskip}{0pt}
  \setlength{\parsep}{0pt}
}
{\end{itemize}}

\newcommand{\problem}[3]
{
  \bigskip
  \noindent
  {\sc #1}
  \begin{indented}
    {\bf Instance:} {#2}
  \end{indented}
  \begin{indented}
    {\bf Question:} {#3}
  \end{indented}
  \bigskip
}

\newcommand{\insX}{\vv{X}}
\newcommand{\insY}{\vv{Y}}
\newcommand{\variables}{{\Gamma_V}}
\newcommand{\body}[1]{\mathit{body}({#1})}
\newcommand{\head}[1]{\mathit{head}({#1})}
\newcommand{\dep}{\Sigma}
\newcommand{\sch}[1]{\mathit{schema}({#1})}
\newcommand{\tup}[1]{\langle {#1} \rangle}
\newcommand{\chase}[2]{\mathit{dchase}({#1},{#2})}
\newcommand{\mods}[2]{\mathit{Mod}({#1},{#2})}

\newcommand{\datalogpm}{$\mbox{Datalog}^\pm$}

\newcommand{\EXP}{{\scshape Exp}}
\newcommand{\PTIME}{\textsc{P}}
\newcommand{\NP}{{\scshape NP}}
\newcommand{\co}{\text{co}}

\newcommand{\AC}[1]{\textsc{$\mbox{AC}_{#1}$}}

\newcommand{\PSPACE}{\textsc{PSpace}}

\let\phi\varphi
\let\epsilon\varepsilon

\newtheorem{theorem}{Theorem}[section]
\newtheorem{definition}[theorem]{Definition}

\title[Reasoning under Disjunctive Existential Rules]
      {The Impact of Disjunction on Reasoning under Existential Rules: Research
      Summary}

\author[M. Morak]
       {MICHAEL MORAK\\
       University of Oxford, Department of Computer Science, OX1 3QD, United
       Kingdom\\
       \email{michael.morak@cs.ox.ac.uk}}

\volume{\textbf{10} (3)}
\jdate{May 2014}
\pubyear{2014}
\pagerange{\pageref{firstpage}--\pageref{lastpage}}
\doi{S0000000000000000}
\jurl{xxxxxx}
\pubdate{15 May 2014}

\pubauthor{Morak}

\begin{document}

\maketitle

\begin{abstract}
  \datalogpm is a Datalog-based language family enhanced with existential
  quantification in rule heads, equalities and negative constraints. Query
  answering over databases with respect to a \datalogpm\ theory is generally
  undecidable, however several syntactic restrictions have been proposed to
  remedy this fact. However, a useful and natural feature however is as of yet
  missing from \datalogpm: The ability to express uncertain knowledge, or
  choices, using disjunction. It is the precise objective of the doctoral thesis
  herein discussed, to investigate the impact on the complexity of query
  answering, of adding disjunction to well-known decidable \datalogpm\
  fragments, namely guarded, sticky and weakly-acyclic \datalogpm\ theories. For
  guarded theories with disjunction, we obtain a strong 2\EXP\ lower bound in
  the combined complexity, even for very restricted formalisms like fixed sets
  of (disjunctive) inclusion dependencies. For sticky theories, the query
  answering problem becomes undecidable, even in the data complexity, and for
  weakly-acyclic query answering we see a reasonable and expected increase in
  complexity.

  \it
  A full version of a paper accepted to be presented at the Doctoral Consortium
  of the 30th International Conference on Logic Programming (ICLP 2014), July
  19-22, Vienna, Austria
\end{abstract}

\begin{keywords}
  Ontological Reasoning, Query Answering, Existential Rules, Logic, TGDs
\end{keywords}

\section{Introduction and Problem Description}
\label{sec:introduction}

For the last thirty years, Datalog (see e.g., \cite{bk:AbiteboulHV95}) has
played an important role as a conceptual query language. Whilst not directly
implemented in mainstream database management systems (DBMS), it did heavily
influence the design of the SQL standard, which now also allows for recursive
statements, as can be expressed in Datalog.

However in recent years it has become increasingly important to add ontological
reasoning capabilities to the existing object-relational querying capabilities
of traditional DBMS: A query is no longer just evaluated over the extensional
relational database, but also over an ontological theory that, using rules and
constraints, describes how to derive new (intensional) knowledge from the
extensional data.  By extending Datalog in such a way that existential
quantification, the first-order logic constant \emph{false} and equalities
between variables are permitted in the rule heads, this behaviour can be
expressed. Recently the \datalogpm\ family of languages has been proposed in
\cite{aaai:CaliGP11}, that defines sensible restrictions on the structure of
such an ontological theory. These restrictions are necessary as, depending on
the structure of the ontological theory, an infinite amount of intensional
knowledge might be derivable, rising the question of decidability of this type
of reasoning. Also, as new values can be invented along the way, the domain can
become infinite.

Despite these obstacles, commercial service providers have already started to
integrate ontological reasoning engines into their database management systems
(see e.g., \cite{online:oraclesemantictech,online:microsoftcsf}), as there are
several applications where such capabilities are desirable, such as data
exchange, ontological reasoning (e.g., reasoning under description logics, or in
the semantic web) and web data extraction.

\paragraph*{Problem Statement.} Given the fact that ontological reasoning is
gaining mainstream acceptance and the fact that, as for example Answer Set
Programming has proven, rule-based languages are well suited for knowledge
representation and reasoning tasks, it is natural to ask how to enrich the
languages we currently have with new, useful constructs. The construct that we
want to focus on here is disjunction. Until now, \datalogpm\ rules only allow us
to express deterministic knowledge. But what about natural statements like
``every person has a parent that is either male or female'' or ``every student
is either an undergraduate or a graduate student''? Such statements are not
captured by existing \datalogpm\ languages. Seeing that disjunctive knowledge is
an important feature in other logical languages like Answer Set Programming or
Description Logics that allows users to intuitively formulate problems by, e.g.,
applying a guess-and-check approach, enriching \datalogpm\ with disjunction is
therefore a logical next step.

The objective of my doctoral studies is thus to introduce the language feature
of disjunction to \datalogpm, and investigate in-depth what the impact of doing
so is w.r.t.\ decidability and complexity of reasoning, focussing on conjunctive
query answering in particular.

\section{Background and Literature Review}
\label{sec:background}

In the following subsections, we give a few basic preliminaries describing
\datalogpm, as well as an overview over the known results in the area.

\subsection{Background}
\label{sub:preliminaries}

In this section the basic notions of conjunctive query evaluation under tuple
generating dependencies (TGDs) are recalled, including a review of the chase
procedure, an important algorithmic tool in the evaluation of queries under
TGDs. Furthermore we briefly introduce the concept of stable models in the logic
programming perspective. We assume that the reader is familiar with first-order
logic as well as basic complexity theory. Good introductions to the former can
be found in e.g.\ \cite{coll:Barwise77} and \cite{bk:Andrews02}, for the latter
we recommend \cite{bk:Papadimitriou94}.

\subsubsection{Conjunctive Queries and the Relational Model}
\label{subsub:conjunctivequeries}

In order to define the semantics of conjunctive queries, we first need to
introduce the relational data model. In the relational data model, the
structure or \emph{schema} ${\cal S}$ of a database and its contents or
\emph{instance} $D$ are distinct objects.

A schema ${\cal S}$ consists of a finite number of \emph{relation symbols} (also
called \emph{predicates}) $r_i$, that is, ${\cal S} = \{r_1,\dots,r_n\}$.

Such a relation symbol $r_i \in {\cal S}$ (for any $i$) consists of a finite
number of \emph{attributes}, such that each attribute has a \emph{domain} of
possible values. We consider here only the case that all predicates have a
common domain $\Gamma\cup\Gamma_N$, where $\Gamma$ is a set of constants and
$\Gamma_N$ is a set of labelled nulls (i.e., distinct null values, each with a
unique name, comparable to skolem constants). The number of attributes of a
relation symbol is called the \emph{arity}, denoted $\mathit{arity}(r_i)$.

A relation $R_i$ for predicate $r_i$ is a set of \emph{tuples} and each tuple is
a mapping of each attribute in $r_i$ to $\Gamma\cup\Gamma_N$. Such a tuple of
$R_i$ is denoted by $r_i(x_1,\dots,x_k)$ (also referred to as an \emph{atom}),
where $k = \mathit{arity}(i_i)$. 

An \emph{instance} $I$ for a schema ${\cal S}$ consists of \emph{relations}
$R_i$ for each $r_i\in{\cal S}$, that is, $D=\{R_1,\dots,R_n\}$. An instance in
which no null values from $\Gamma_N$ appear is referred to as a \emph{database},
usually denoted $D$. Note that, when viewed as a first-order theory, we may
simply interpret an instance as a conjunction of atoms.

A \emph{conjunctive query} $q$ over a database schema ${\cal S}$ is an assertion
of the form $$q(\vv{X})\leftarrow\exists\vv{Y}\phi(\vv{X},\vv{Y})$$ where
$\vv{X}$ and $\vv{Y}$ are vectors of (first-order logic) variables, $q(\vv{X})$
is called the \emph{head}, $\mathit{dimension}(\vv{X})$ is called the
\emph{arity} of $q$ and $\phi(\vv{X},\vv{Y})$ is called the \emph{body}, where
$\phi(\vv{X},\vv{Y})$ is a first-order formula consisting of a conjunction of
atoms of the form $r_i(t_1,\dots,t_k)$ and equalities of the form $t_1=t_2$,
where $r_i$ is a predicate of ${\cal S}$ with arity $k$ and each $t_i$ is either
a constant from $\Gamma$ or a (first-order logic) variable. If the arity is $0$
then $q$ is called a \emph{boolean conjunctive query}.

With every database $D=\{R_1,\dots,R_n\}$ over a schema ${\cal S}$, we can now
associate a finite first-order structure $M_D=(U,R_1,\dots,R_n)$ with universe
$U=\Gamma$. The evaluation of a conjunctive query $q$ then comes down to
checking satisfiability in first-order logic as follows: $q$ has an answer over
$D$, denoted $D\models q$, if and only if the set $\{\langle
a_1,\dots,a_k\rangle\mid M_D\models q(a_1,\dots,a_k)\}$ is non-empty, with $a_i
\in \Gamma$. This set is also called the set of \emph{answers} to $q$ over $D$,
where $k$ is the arity of $q$.

\subsubsection{Dependencies}
\label{subsub:dependencies}

For reasoning tasks over databases, the need arises to express how new
(\emph{intensional}) knowledge can be derived from the data that is stored in
the database (called the \emph{extensional data}). An established way to do this
is to introduce a set $\Sigma$ of rules that describe the relation between
intensional and extensional data. In this case for a database $D$, the logical
theory $D\cup\Sigma$, i.e., the conjunction of the facts in the database with
all the rules in $\Sigma$, is taken as a basis for conjunctive query evaluation.

Rules in $\Sigma$ over a schema ${\cal S}$ are of either one of the following
two forms:
\begin{align}
  \label{eq:tgd}\forall\vv{X}
    (\phi(\vv{X})&\rightarrow\exists\vv{Y}\psi(\vv{X},\vv{Y}))\\
  \label{eq:egd}\forall\vv{X}(\phi(\vv{X})&\rightarrow X_i = X_j)
\end{align}
where rules of the form of (\ref{eq:tgd}) are referred to as tuple generating
dependencies (TGDs) and of (\ref{eq:egd}) as equality generating dependencies
(EGDs), with $\phi$ and $\psi$ being conjunctions of predicates from ${\cal S}$
(also called \emph{atoms}) and $X_i$ and $X_j$ are the $i$-th and $j$-th
position in vector $\vv{X}$. $\phi$ is also referred to as the body of the
dependency and $\psi$ or $X_i = X_j$ as the head. TGDs where $\psi = \bot$ are
called \emph{negative constraints}. For brevity, we will omit the universal
quantifiers in front of TGDs and EGDs, and replace conjunctions in the body by
commas.

Given an instance $I$, it is said to be satisfying a dependency
$\sigma\in\Sigma$, that is, $I\models\sigma$, if the first-order sentence formed
by a conjunction of the facts in $I$ and $\sigma$ is satisfiable. By extension,
$I$ satisfies $\Sigma$ ($I \models \Sigma$) iff it satisfies every
$\sigma\in\Sigma$.

The models of a database $D$ over a schema ${\cal S}$ with respect to $\Sigma$,
denoted $\mathit{Mod}(D,\Sigma)$, are all instances $M$ that satisfy
$D\cup\Sigma$ (i.e., $I \supseteq D$ and $I \models \Sigma$). When answering
conjunctive queries we use the certain answer semantics, i.e., we consider the
query to be true only iff it is true under every model. The set of answers for a
conjunctive query $q$, denoted $\mathit{ans}(q,D,\Sigma)$, thus equals the set
$$\{\langle a_1,\dots,a_k\rangle \mid \forall M\in\mathit{Mod}(D,\Sigma):
M\models q(a_1,\dots,a_k)\}$$

For complexity analysis we focus on the decision version of this problem. This
is the central problem when analyzing the complexity of databases, tuple and
equality generating dependencies and therefore \datalogpm\ complexity issues.
Below it is formulated for boolean conjunctive queries, which we will focus on
in this work:

\problem{BCQ-Answering}
{
  $\langle q,D,\Sigma\rangle$: $q$ a boolean conjunctive query, $D$ a
  database and $\Sigma$ a set of dependencies
}
{
  $D\cup\Sigma\models q$?
}

Usually when dealing with query evaluation over databases the \emph{data
complexity} and the \emph{combined complexity} are of interest. In this paper
we follow the approach of \cite{stoc:Vardi82} where for the former everything
except the database $D$ is considered fixed, i.e., the only input is the
database. For the latter, the database $D$, $\Sigma$ and the query itself form
the input. 

Unfortunately, in general it holds that {\sc BCQ-Answering} is undecidable under
unrestricted sets of TGDs, as has been shown in \cite{icalp:BeeriV81}. In
\cite{CaliGK13,tr:BagetLM09,ai:BagetLMS11} it has further been shown that even
singleton sets of TGDs cause query answering to become undecidable.

These results clearly show that restrictions must be placed on the structure of
$\Sigma$ to ensure decidability. This is a non-trivial problem, as simple
restrictions, like limiting the number of TGDs, are not enough.

\subsubsection{The Chase}
\label{subsub:thechase}

One of the fundamental tools to algorithmically check implication of
dependencies is the \emph{chase procedure}, introduced in
\cite{tods:MaierMS79}, which was later adapted for checking query containment
in \cite{jcss:JohnsonK84}, in the setting of databases with tuple and equality
generating dependencies, or, more specifically, in the setting of databases
with inclusion and functional dependencies. The chase algorithm tries to
extend a given database instance in such a way that every TGD and EGD becomes
satisfied. This is done by exhaustively (i.e., until a fix-point is reached)
applying the \emph{chase step}:

\begin{definition}
  Let $D$ be a database and $\Sigma$ be a set of dependencies. A
  \emph{chase step} is defined as follows:

  \medskip
  {\bf TGDs.} Let $\Sigma$ contain a TGD $\phi(\vv{X}) \rightarrow \exists
  \vv{Y} \psi(\vv{X},\vv{Y})$, such that
  \begin{packeditems}
    \item $D \models \phi(\vv{a})$ for some assignment $\vv{a}$ to $\vv{X}$,
      and
    \item $D \nvDash \exists \vv{Y} \psi(\vv{a},\vv{Y})$.
  \end{packeditems}
  Then extend $D$ with facts $\psi(\vv{a},\vv{y})$, where the
  elements of $\vv{y}$ are fresh labelled nulls (i.e., values from $\Gamma_N$
  that have not been in use in $D$ up to that point.

  \medskip
  {\bf EGDs.} Let $\Sigma$ contain an EGD $\phi(\vv{X}) \rightarrow X_i =
  X_j$, such that
  \begin{packeditems}
    \item $D \models \phi(\vv{a})$ for some assignment $\vv{a}$ to $\vv{X}$,
      and
    \item $a_i \neq a_j$
  \end{packeditems}
  If $a_i$ is a labelled null, then replace every occurrence of $a_i$ with
  $a_j$ or vice-versa if $a_j$ is a labelled null. If $a_i$ and $a_j$ are
  distinct constants, end the chase with failure.
\end{definition}

\begin{definition}
  The \emph{chase expansion} of a database instance $D$ with respect to a set of
  dependencies $\Sigma$ is a sequence $D_0, D_1, \dots, D_m$, such that $D_0 =
  D$ and for $i \geq 0$, $D_{i+1}$ is obtained from $D_i$ by applying a chase
  step. After exhaustively applying such chase steps, we obtain $D_m$, also
  denoted $\mathit{chase}(D, \Sigma)$.
\end{definition}

The chase can have three different outcomes: Failure, non-terminating success or
terminating success. In case of success the resulting instance $D_m$ satisfies
all dependencies in $\Sigma$. Note that if the chase does not terminate, $m =
\inf$ and the size of $D_m$ is infinite.

We assume that the chase is fair, i.e., we exclude the possibility of a
degenerated chase expansion by assuming that the chase expansion is constructed
level by level, and after each application of a TGD, all applicable EGDs are
applied. This ensures that every TGD that can be applied, is applied, and
therefore we exclude the case that only a single infinite path in the chase
expansion is ever expanded when in case the chase is infinite.

\paragraph*{Query Answering and the Chase} In case the chase succeeds, it
computes a \emph{universal solution} for $\langle D, \Sigma\rangle$. Every model
$M \in \mathit{Mod}(D,\Sigma)$ can then be obtained by appropriate instantiation
of labelled nulls in $chase(D,\Sigma)$ (i.e., for every model $M$, there exists
a homomorphism mapping the universal solution to $M$; cf.\ \cite{DeNR08}). Using
this property, the chase expansion of a database $D$, with respect to a set of
dependencies $\Sigma$, can be used for answering conjunctive queries, as the
following theorem shows:

\begin{theorem}[\cite{DeNR08}]
  Given a boolean conjunctive query $q$ over a schema ${\cal S}$, a database $D$
  of ${\cal S}$ and a set of dependencies $\Sigma$ over ${\cal S}$, then in
  cases where the chase does not fail, it holds that $D \cup \Sigma \models q$
  if and only if $\mathit{chase}(D,\Sigma) \models q$.
\end{theorem}

In case the chase fails, query answering is trivial: As there is no model, every
boolean conjunctive query clearly is entailed by $D\cup\Sigma$ (cf.\ the
definition of certain answers in section \ref{subsub:dependencies}).

\subsection{Literature Review}
\label{sub:decidabilityandcomplexity}

In this section we discuss the different kinds of restrictions known to ensure
decidability of query answering under sets of TGDs. The decidable classes of
TGDs discussed below are defined by syntactic properties that either apply to
single TGDs (local syntactic conditions) or to the set of all TGDs (global
syntactic conditions). These properties can be checked in finite time using
appropriate algorithms. Each subsection deals with a known syntactic condition
that ensures decidability of query answering.

\paragraph{Inclusion Dependencies} Inclusion dependencies (IDs), one of the
simplest forms of dependencies, allow one to express that certain values
occurring in a specific position in one relation, must also occur at (or be
included in) a specific position in another relation.  This allows for TGDs that
consist of one body and head atom only, and no variable may occur twice in the
head or the body. The following is an example of an inclusion dependency,
expressing that every student is a person:
$$\mathit{student}(X, Y) \to \exists Z \, \mathit{person}(X, Z)$$
The query answering problem was shown to be decidable, and in fact in \AC{0}\
(resp.\ \PSPACE) in the data (resp.\ combined) complexity.

\paragraph{Linear Tuple Generating Dependencies} This class is similar to IDs in
that it allows for TGDs with only a single body atom, but generalizes them,
because it allows repetition of variables in the body or head (e.g., the TGD
$r(X,Y,X) \rightarrow s(X,Y)$ is a linear TGD but not an ID).

Sets of linear TGDs enjoy the so-called \emph{bounded derivation-depth property
(BDDP)}, which roughly implies that only a finite initial part of the chase is
required for query answering, thus ensuring decidability. As with inclusion
dependencies, first-order rewritability (i.e., rewriting $q$ and $\Sigma$ into a
first-order query $q_\Sigma$, such that $D \cup \Sigma \models q$ iff $D \models
q_\Sigma$) is thus possible (cf. \cite{pods:CaliGL09,amw:CaliGP10}). Therefore
we get decidability, and query answering is in \AC{0} in the data complexity.
Regarding combined complexity, results from inclusion dependencies carry over to
linear TGDs, resulting in the \PSPACE-completeness for query answering in the
general case and \NP-completeness in case of a fixed set of TGDs.

\paragraph{Guarded Tuple Generating Dependencies} In \cite{CaliGK13}, linear
TGDs are extended to so-called \emph{guarded} TGDs, that have a body atom that
contains all variables occurring in the body, i.e., all universally quantified
variables. This atom is called the \emph{guard}. If there are multiple such
atoms, the leftmost is taken as the guard. An example of a guarded TGD that says
that if students are in their first semester, they have a tutor, is as follows.
Note that it is not linear as it has multiple atoms in the body.
$$\mathit{student}(X, Y), \mathit{firstsemester}(X) \to \exists Z \,
\mathit{tutor}(X, Z)$$
Linear TGDs and inclusion dependencies are trivially guarded, as they only have
exactly one body atom. However, guarded TGDs are not first-order rewritable.
This is shown by creating a database, query and a set of guarded TGDs in such a
way that answering the query requires the computation of the transitive closure
over a relation in the database. It is well known that this property cannot be
expressed in a finite first-order query, and we cannot obtain decidability
thusly. However, it can be shown that the universal model constructed by the
chase, albeit possibly infinite, is of finite treewidth (i.e., it is tree-like
and cannot be arbitrarily cyclic). From Courcelle's famous Theorem (cf.\
\cite{Cour90}), which states that evaluating first-order sentences over
structures of finite treewidth is decidable, we derive decidability for query
answering under sets of guarded TGDs.

The complexity of query answering under guarded TGDs was investigated in
\cite{pods:CaliGL09}, where it was established that, whenever a query is
actually entailed by a database and a set of guarded TGDs, then all atoms needed
to answer the query are derived in a finite, initial portion of the chase when
restricted only to guards and atoms derived from them, whereby the size of this
portion depends only on the query and the set of TGDs. Therefore, constructing
this part of the chase and evaluating a boolean conjunctive query over it is
enough to compute the answer. It is shown that this can be done in polynomial
time in the data complexity, whereby \PTIME-membership follows. Hardness for
\PTIME\ was shown in \cite{csur:DantsinEGV01} by reduction to the implication
problem for propositional logic programs.

The combined complexity is investigated in \cite{CaliGK13}, proving the
2\EXP-completeness for the general case and \EXP-completeness in case of fixed
arity. Also membership in \NP\ was shown in case where the set of TGDs is fixed.
\NP-hardness follows from results in \cite{stoc:ChandraM77}, which show that
\NP-hardness holds even for the empty set of TGDs.

\paragraph*{Weakly-Guarded Sets of TGDs} In \cite{CaliGK13}, guarded TGDs were
extended to \emph{weakly-guarded sets of TGDs}. Every TGD in such sets must have
an atom in its body that contains all the variables where a null value may
appear during the chase. The leftmost such atom is called the \emph{weak-guard}.
This class is the first class discussed here that is based on a global property.
It is easy to see that, as guarded TGDs contain a body atom with all universally
quantified variables, they are trivially weakly-guarded, as the guard is also a
weak-guard.

It is implicit in \cite{CaliGK13} that it can be verified in polynomial time
whether a set of TGDs is weakly-guarded or not: For a schema ${\cal S}$ we first
need to compute all the positions for each predicate where a null value can
occur during the chase with respect to a set of TGDs $\Sigma$. These positions
are called \emph{affected} and computing them has been shown to be possible in
polynomial time. Then we have to check for each TGD in $\Sigma$ whether it
contains a weak-guard, which, knowing the affected positions, is also possible
in polynomial time. 

It is then shown that weakly-guarded sets of TGDs enjoy the same favorable
property as guarded TGDs, namely, the chase has finite treewidth.  Given this
fact, decidability of query answering is established as before. Regarding the
complexity, in general the problem is 2\EXP-complete, \EXP-complete if the arity
is fixed or the set of TGDs fixed, and it remains \EXP-complete even if only the
database is considered as input (data complexity).

\paragraph{Weakly-Acyclic Sets of Tuple Generating Dependencies} The notion of
\emph{Weak Acyclicity} was established in the landmark paper
\cite{tcs:FaginKMP05} as a syntactic condition to guarantee termination of the
chase procedure. For this we first need to define the notion of a dependency
graph.

A dependency graph $G=(V,E)$ is constructed as follows: $V$ is the set of
attributes of all the relations occurring in $\Sigma$. We will denote the $i$th
attribute of some relation $r$ by $r[i]$. For each TGD $\sigma = \phi(\vv{X})
\rightarrow \exists \vv{Y} \psi(\vv{X}, \vv{Y})$ and each variable $X \in
\vv{X}$ shared between the relation attributes $r[i]$ in $\phi$ and $s[j]$ in
$\psi$, we add an edge $(r[i],s[j])$ to $E$. We add a special edge $(r[i],p[k])$
to $E$ for each attribute $p[k]$ in $\psi$ occupied by a variable $Y \in
\vv{Y}$, and each attribute $r[i]$ occurring in the body of $\sigma$.

A set $\Sigma$ of TGDs is called \emph{weakly-acyclic} if its dependency graph
contains no cycles through special edges. The definition of weak acyclicity is a
global property and can be decided in \PTIME, as the construction of the
dependency graph and the cycle-check through a special edge are both feasible in
\PTIME.

In \cite{tcs:FaginKMP05} it was shown that for weakly-acyclic sets of TGDs the
chase always terminates. This is ensured by the fact that when cycles through
special edges in the dependency graph are forbidden, no new null values can be
added in a later chase step because of a null value added in an earlier chase
step. Therefore we trivially get decidability: Simply compute the (finite)
chase, and then answer the query on the obtained finite model. 

Regarding complexity (cf. \cite{CaliGK13,pods:KolaitisPT06}), in general the
problem of {\sc BCQ-Answering} is 2\EXP-complete for weakly-acyclic sets of
TGDs. When the set of TGDs is fixed, the {\sc BCQ-Answering} problem is known to
be \NP-complete. \PTIME-completeness holds for the data complexity, following
from the complexity of the fact inference problem for fixed Datalog programs
(see \cite{csur:DantsinEGV01}).

\paragraph{Sticky Sets of Tuple Generating Dependencies} A recent addition to
the set of syntactic conditions that ensure decidability and favourable
complexity of conjunctive query evaluation is the paradigm of \emph{stickiness},
introduced in \cite{CaliGP12}. A survey of sticky classes can be found in
\cite{lics:CaliGLMP10}. The class of \emph{sticky sets of TGDs} is defined as
follows: In a first step, a variable marking of all TGDs in a set $\Sigma$ is
computed by a procedure called {\sf SMarking}. This is a two-step procedure:

\begin{packedenum}
  \item \emph{Initial marking:} For each $\sigma \in \Sigma$, if there exists
    a variable $V$ in the body of $\sigma$ and an atom without this variable
    exists in the head of $\sigma$, mark each occurrence of $V$ in the body.
  \item \emph{Propagation step:} Until a fixpoint is reached, consider any
    pair $\langle\sigma_1,\sigma_2\rangle\in\Sigma\times\Sigma$. If a
    universally quantified variable $V$ occurs in $\mathit{head}(\sigma_1)$ at
    positions $\pi_1,\dots,\pi_m$ for $m \geq 1$ and an atom in
    $\mathit{body}(\sigma_2)$ exists where at each of these same positions a
    marked variable occurs, then mark each occurrence of $V$ in
    $\mathit{body}(\sigma_1)$.
\end{packedenum}

\begin{definition}[\cite{CaliGP12}]
  A set $\Sigma$ of TGDs is called \emph{sticky} if and only if there is no TGD
  in ${\sf SMarking}(\Sigma)$ such that a marked variable occurs in its body
  more than once.
\end{definition}

The property of stickiness is incomparable to guardedness and weak acyclicity
but strictly generalizes inclusion dependencies. In comparison to other
discussed syntactic classes of TGDs, sticky sets of TGDs allow for a mildly
restricted way to express joins. The following is an example of a sticky
(singleton) set of TGDs, expressing the join between two tables, department and
employee, to get a combined table of departments and their heads:
$$\mathit{department}(X, Y), \mathit{employee}(Y, Z) \to \mathit{headofdept}(X,
Y, Z)$$
Note that the above TGD is not weakly-guarded.

\section{Goal and Current Status of the Research}
\label{sec:goalstatus}

The goal, as already discussed in the introduction, is to introduce disjunction
into \datalogpm\ and investigate the impact of doing so on the decidability and
complexity of query answering. We thus extend the definition of a TGD to allow
for disjunction as follows:

A \emph{disjunctive tuple-generating dependency (DTGD)} $\sigma$
is a first-order formula $ \forall \insX \, \varphi(\insX)\, \rightarrow\,
\bigvee_{i=1}^{n} \exists \insY \, \psi_i(\insX,\insY), $ where $n \geqslant 1$,
$\insX \cup \insY \subset \variables$, and $\varphi,\psi_1,\ldots,\psi_n$ are
conjunctions of atoms; $\varphi$ is the $\emph{body}$ of $\sigma$, denoted
$\body{\sigma}$, while $\bigvee_{i=1}^{n} \psi_i$ is the \emph{head} of
$\sigma$, denoted $\head{\sigma}$. If $n=1$, then $\sigma$ is a
\emph{tuple-generating dependency (TGD)}. Given a set $\dep$ of DTGDs,
$\sch{\dep}$ is the set of predicates occurring in $\dep$.

We employ the \emph{disjunctive chase} introduced in~\cite{DeTa03} in order to
answer queries. It is an extension of the chase procedure described in Section
\ref{sub:preliminaries}.
Consider an instance $I$, and a DTGD $\sigma = \varphi(\insX) \rightarrow
\bigvee_{i=1}^{n} \exists \insY\,\psi_i(\insX,\insY)$.  We say that $\sigma$ is
\emph{applicable} to $I$ if there exists a homomorphism $h$ (i.e., a
substitution of labelled nulls to either constants or other labelled nulls) such
that $h(\varphi(\insX)) \subseteq I$, but there is no $i \in \{1,\ldots,n\}$ and
a homomorphism $h' \supseteq h$ such that $h'(\psi_i(\insX,\insY)) \subseteq I$.
The result of applying $\sigma$ to $I$ with $h$ is the set $\{I_1,\ldots,I_n\}$,
where $I_i = I \cup h'(\psi_i(\insX,\insY))$, for each $i \in \{1,\ldots,n\}$,
and $h' \supseteq h$ is such that $h'(Y)$ is a ``fresh'' labelled null not
occurring in $I$, for each $Y \in \insY$. For such an application of a DTGD,
which defines a single DTGD \emph{chase step}, we write $I \tup{\sigma,h}
\{I_1,\ldots,I_n\}$.

A \emph{disjunctive chase tree} of a database $D$ and a set $\dep$ of DTGDs is a
(possibly infinite) tree such that the root is $D$, and for every node $I$,
assuming that $\{I_1,\ldots,I_n\}$ are the children of $I$, there exists $\sigma
\in \dep$ and a homomorphism $h$ such that $I \tup{\sigma,h}
\{I_1,\ldots,I_n\}$.
The disjunctive chase algorithm for $D$ and $\dep$ consists of an exhaustive
application of DTGD chase steps in a fair fashion, which leads to a disjunctive
chase tree $T$ of $D$ and $\dep$; we denote by $\chase{D}{\dep}$ the set
$\{I~|~I \textrm{~is~a~leaf~of~} T\}$. Note that each leaf of $T$ is
well-defined as the least fixpoint of a monotonic operator.
By construction, each instance of $\chase{D}{\dep}$ is a model of $D$ and
$\dep$. Interestingly, $\chase{D}{\dep}$ is a \emph{universal set model} of $D$
and $\dep$, i.e., for each $M \in \mods{D}{\dep}$, there exists $I \in
\chase{D}{\dep}$ and a homomorphism $h_I$ such that $h_I(I) \subseteq
M$~\cite{DeNR08}. This implies that w.r.t.\ certain answers, given a query $q$,
$D \cup \dep \models q$ iff $I \models q$, for each $I \in \chase{D}{\dep}$.

\paragraph{Current Status.} Currently we have investigated and obtained results
for all the decidable classes of TGDs. For the guarded-based classes, adding
disjunction does not make the problem of query answering undecidable. However it
does in certain cases increase the complexity of the problem by a significant
amount. For the guarded-based classes of TGDs (i.e., IDs, linear, guarded and
weakly-guarded), we have established all relevant complexity results when
extending them to DTGDs.

In case of sticky TGDs, when adding disjunction the problem of query answering
becomes undecidable. This was a very surprising result, given the fact that the
complexity of query answering under sticky sets of TGDs is lower than under
guarded TGDs.

In case of weakly-acyclic TGDs, data complexity results have been obtained, as
well as certain lower bounds in the combined complexity, however a matching
upper bound is still missing here. Decidability is assured in any case, because
the disjunctive chase terminates, which follows from the definition of weak
acyclicity.

\section{Preliminary Results}
\label{sec:results}

One classical work on disjunction in ontologies is \cite{kr:CalvaneseGLLR06},
which immediately gives us \co\NP-hardness for conjunctive query answering over
disjunctive ontologies, even if the query is fixed, and the ontology consists of
a fixed, single rule of the form $a(X) \to b(X) \vee c(X)$. Without restricting
the query language, there is thus no hope to get tractability results. However,
for atomic queries, where the query consists only of a single atom, there are
tractable data complexity cases to be found.

\paragraph{Arbitrary queries.} In \cite{BourhisMP13}, we have investigated the
complexity picture for answering arbitrary queries. The main results are as
follows:

\begin{packeditems} 
  \item 2\EXP-completeness whenever the query is non fixed.
    This is shown by simulating a B\"uchi tree automaton, and it even holds for
    fixed sets of Disjunctive Inclusion Dependencies (DIDs) of arity at most
    three, or of non-fixed sets of the same with arity at most two.

  \item \co\NP-completeness in the data complexity for query answering under
    DIDs up to guarded DTGDs.

  \item \EXP-completeness in the data complexity for query answering under
    weakly-guarded sets of DTGDs.
\end{packeditems}

In case of (non-disjunctive, classical) TGDs, complexity results coincide in the
data complexity, but vary from \co\NP-completeness to 2\EXP-completeness for
fixed sets of IDs to weakly-guarded sets of TGDs. It is thus interesting to note
that adding disjunction to expressive languages doesn't change the complexity in
this case, but there is a high cost to add it to less expressive languages.

\paragraph{Atomic queries.} In \cite{GottlobMMP12}, we have investigated the
complexity of answering single-atom queries. Here the complexity results vary
considerably:

\begin{packeditems}
  \item 2\EXP-completeness in the combined complexity for guarded DTGDs.

  \item \EXP-completeness in the combined complexity for linear DTGDs.

  \item \co\NP-completeness in the data complexity for guarded DTGDs.

  \item Membership in \AC{0} in the data complexity for linear DTGDs.
\end{packeditems}

In the case of atomic queries we do have a number of tractability results to
offer, especially the highly parallelizable data complexity of \AC{0} in case of
atomic query answering over sets of linear DTGDs (which captures the class of
DIDs). For guarded DTGDs, most of the results follow directly from expressive
fragments of first-order logic (the Guarded Fragment
\cite{BaGO10,jsyml:Gradel99}, and Guarded-Negation First-Order Logic
\cite{BaCS11,BaCO12}). For linear, we develop novel machinery to obtain our
respective bounds.

\section{Open Issues and Expected Achievements}
\label{sec:openissues}

In addition to the published results, we would like to find answers to the
following questions: What is the complexity of query answering under sets of
\begin{packedenum}
  \item guarded-based DTGDs in case where the query is acyclic or of bounded
    (hyper)treewidth?
  \item weakly-acyclic sets of DTGDs?
  \item sticky sets of DTGDs?
\end{packedenum}

Regarding the first item, we have already managed to obtain all the relevant
results. In fact, for bounded (hyper)treewidth, the complexity table coincides
with that of arbitrary queries. For acyclic queries, there are drops in
complexity corresponding to the expressivity of the language considered. Papers
containing these results have been submitted to this year's MFCS conference and
DL workshop. It is our plan to subsequently publish these results, in addition
to some extended work on arbitrary and atomic queries in a comprehensive journal
paper, treating all the guarded-based classes of DTGDs, in the course of 2014.

Regarding weakly-acyclic, we already have answers to the complexity questions
for data complexity and the cases of fixed sets and fixed arities. However, we
are still missing the combined complexity results. Before submission of my
thesis, we plan to close these open complexity questions as well.

Lastly, for sticky DTGDs, we have an undecidability proof, which is somewhat
surprising as query answering under sticky TGDs is easier in terms of complexity
than it is for guarded TGDs, yet the addition of disjunction doesn't cause a
complexity increase in the latter. We have therefore focused on extending
guarded DTGDs with cross-products (a form of join allowed in sticky TGDs). This
again yields undecidability, however it becomes decidable if restricted to arity
at most two, where binary predicates can never participate in a disjunction. For
this case we are working on obtaining the relevant complexity results.

\bibliographystyle{acmtrans}
\bibliography{iclpdc2014,shorttitles,references}

\end{document}